\documentclass[useAMS,usenatbib,usegraphicx]{mn2e}

\usepackage{times}
\usepackage{amsfonts,amssymb}
\usepackage{color}

\title[Feedback \& the Abundance and Clustering of HI]
      {On the role of feedback in shaping the cosmic abundance and 
        clustering of neutral atomic hydrogen in galaxies}
      \author[Kim et al.]
             { \parbox{\textwidth}{Han-Seek~Kim$^{1}$\thanks{E-mail:\texttt{hansikk@unimelb.edu.au}}, 
                 C.~Power$^{2,4}$\thanks{E-mail:\texttt{chris.power@icrar.org}}, C. M.~Baugh$^3$, J. S. B.~Wyithe$^{1,4}$, 
                 C. G.~Lacey$^{3}$, \\
                 C. D. P.~Lagos$^{3}$ \& C. S. Frenk$^{3}$}
               \vspace{0.4cm}\\
                \parbox{\textwidth}{$^1$School of Physics, The University of 
                 Melbourne, Parkville, VIC 3010, Australia\\
                 $^2$International Centre for Radio Astronomy Research, 
                 University of Western Australia, 35 Stirling Highway, 
                 Crawley, WA 6009, Australia\\
                 $^3$Institute for Computational Cosmology, 
                 Department of Physics, University of Durham, 
                 South Road, Durham DH1 3LE, UK\\
                 $^4$ ARC Centre of Excellence for All-Sky Astrophysics (CAASTRO) 
                 }}
\begin{document}

\date{}

\pagerange{\pageref{firstpage}--\pageref{lastpage}} \pubyear{}

\maketitle

\label{firstpage}

\begin{abstract}
  We investigate the impact of feedback -- from supernovae (SNe), active 
  galactic nuclei (AGN) and a photo-ionizing background at high redshifts 
  -- on the neutral atomic hydrogen (HI) mass function, the $b_{\rm J}$ band 
  luminosity function, and the spatial clustering of these galaxies at $z$=0. 
  We use a version of the semi-analytical galaxy formation model 
  {\texttt{GALFORM}} that calculates self-consistently the amount of HI in a 
  galaxy as a function of cosmic time and links its star formation rate to its 
  mass of molecular hydrogen (H$_2$). We find that a systematic increase or 
  decrease in the strength of SNe feedback leads to a systematic decrease or 
  increase in the amplitudes of the luminosity and HI mass functions, but has
  little influence on their overall shapes. Varying the strength of AGN 
  feedback influences only the numbers of the brightest or most HI massive 
  galaxies, while the impact of varying the strength of photo-ionization 
  feedback is restricted to changing the numbers of the faintest or least HI 
  massive galaxies.
  {Our results suggest that the HI mass function is a more
    sensitive probe of the consequences of cosmological reionization for galaxy formation
    than the luminosity function.
  We find that increasing the strength of any of the modes of feedback acts to
  weaken the clustering strength of galaxies, regardless of their HI-richness. In contrast,
  weaker AGN feedback has little effect on the clustering strength whereas weaker SNe 
  feedback increases the clustering strength of HI-poor galaxies more strongly than HI-rich 
  galaxies. These results indicate that forthcoming HI surveys on next generation radio telescopes such 
  as the Square Kilometre Array and its pathfinders will be exploited most fruitfully
  as part of multiwavelength survey campaigns.}

  \end{abstract}
  

\section{Introduction}

Neutral hydrogen, both atomic (HI) and molecular (H$_2$), plays a 
fundamental role in galaxy formation, principally as the raw material 
from which stars are made. The amount of neutral hydrogen in a galaxy 
at any given time reflects the complex interplay between processes 
that either deplete it, such as supernovae, or replenish it, such as gas 
cooling from hot atmospheres surrounding galaxies and by mergers with other 
gas-rich galaxies. By quantifying the properties of HI in galaxies and noting 
how these properties vary with galaxy morphology and environment, we can glean 
insights into the physics of galaxy formation and test predictions of 
theoretical models. 

Most of what we know about the HI properties of galaxies comes from 
surveys of the nearby Universe ($z\lesssim 0.05$) such as HIPASS 
\citep[HI Parkes All-Sky Survey; see][]{meyer.etal.2004} 
and ALFALFA 
\citep[Arecibo Legacy Fast ALFA Survey; see][]{giovanelli.etal.2005}.
Such surveys have revealed that HI-rich galaxies tend to be late-type 
\citep[e.g.][]{kilborn.etal.2002,evoli.etal.2011}, but it is common for 
early-type field galaxies to host HI \citep[e.g.][]{serra.etal.2012}; 
that the HI mass function is well described by a Schechter function 
\citep[e.g.][]{zwaan.etal.2003,martin.etal.2010} but its shape depends 
on environment 
\citep[e.g.][]{zwaan.etal.2005,springob.etal.2005,kilborn.etal.2009}; 
and that HI-rich galaxies are among the most weakly clustered galaxies 
known \citep[e.g.][]{meyer.etal.2007,basilakos.etal.2007,passmoor.etal.2011}. 
However, knowledge of the HI properties of galaxies will improve many-fold over 
the coming decade with the advent of next generation HI galaxy surveys on ASKAP 
\citep[cf.][]{askap.science.2008}, MeerKAT \citep[cf.][]{meerkat.2007}) 
and ultimately the SKA itself \citep[e.g.][]{baugh.etal.2004,blake.etal.2004,
power.etal.2010,abdalla.etal.2010,kim.etal.2011}.

Future galaxy surveys on the SKA and its pathfinders are expected to
revolutionize our view of the HI Universe and so it is timely to ask precisely 
what these surveys can teach us about the physical processes that drive 
galaxy formation. In this paper we focus on feedback -- from stars in the 
form of winds driven by supernovae (SNe), from accreting super-massive black 
holes in the form of active galactic nuclei (AGN) heating, and from a 
photo-ionizing background in the early Universe -- and evaluate how the 
global properties of HI in galaxies such as the HI mass function,
which quantifies the number density of galaxies of a given HI mass, and the
2-point correlation function and halo occupation distributions 
\citep[HODs; e.g.][]{kim.etal.2011}, which quantify spatial clustering, are 
shaped by different sources of feedback. 

In galaxy formation models, the strengths of these processes and the interplay 
between them have been traditionally set by examining the predictions of the galaxy 
luminosity function in the $b_{\rm J}$ band, so we will also study how this statistic 
changes. To do this, we use the version of the semi-analytical galaxy formation model 
{\texttt{GALFORM}} of \citet{cole.etal.2000} as it has been developed by
 \citet{lagos.etal.2011a}; this calculates self-consistently the HI properties 
of galaxies by splitting their interstellar media (ISM) into HI and H$_2$ 
phases using the empirical relations of \citet{blitz.2006} and 
\citet{leroy.etal.2008}, and links star formation in galaxies not to their 
cold gas masses, as assumed in previous models 
\citep[cf.][]{cole.etal.2000,baugh.2006}, but to their H$_2$ masses, as 
suggested by recent observations \citep[e.g.][]{bigiel.etal.2008}. As shown 
in \citet{lagos.etal.2011b}, this model reproduces the observed HI mass 
function at $z$=0, accurately reproducing its amplitude and shape at 
intermediate and low HI masses.

This is particularly interesting because we expect feedback to influence the 
global properties of galaxies selected either by their stellar mass or light.
Previous studies have shown that SNe are pivotal in fixing the amplitude and
slope of the luminosity function 
\citep[e.g.][]{cole.etal.2000,benson.etal.2003a} while AGN heating 
suppresses the formation of massive galaxies and dictates the 
form of the bright end \citep[cf.][]{bower.etal.2006,croton.etal.2006}. 
In contrast, the influence of SNe and AGN on the HIMF appears more 
subtle. For example, \citet{power.etal.2010} found that galaxy 
formation models that included or excluded AGN heating (such as 
\citealt{delucia.blaizot.2007} and \citealt{baugh.etal.2005} 
respectively) predict HI mass functions that reproduce equally well the 
observed $z\!\simeq\!0$ data. The HI mass function offers the possibility of 
placing stronger constraints on the strength of feedback in semi-analytical 
models, which previously used only the optical properties of galaxies 
\citep[e.g.][]{benson.etal.2003a,bower.etal.2010}. At the same time, the 
variation of clustering strength with galaxy properties provides us with 
important clues about the physics of galaxy formation. Any discrepancy between 
observational measurements of clustering and the predictions of galaxy 
formation models indicates the need to improve the models, either by refining 
the modelling of the physical processes included or by considering the addition 
of further effects \citep[see, for example,][]{kim.etal.2009}.

The structure of the paper is as follows. In \S\ref{sec:model}, we provide
a brief overview of the \citet{lagos.etal.2011a} model and describe how the
different modes of feedback (SNe, AGN and photo-ionization) are implemented
in {\texttt{GALFORM}}. In \S\ref{sec:himf} we show how the different forms of
feedback influence both the galaxy luminosity function in $b_{\rm J}$ band and
the HI mass function, {and we deduce using likelihood maximization precisely 
what information we glean from them.} In \S\ref{sec:clustering} we investigate 
how the spatial clustering of HI-poor and HI-rich galaxies are influenced by 
feedback by inspecting the 2-point correlation function and HODs. Finally, in 
\S\ref{sec:summary}, we summarize our results and discuss their implications 
for testing the modelling of feedback with forthcoming HI surveys.

\section{Modelling} 
\label{sec:model}

In this section we first briefly give an overview of the version of 
{\texttt{GALFORM}} used in this paper (\S~\ref{GFM}). Because the focus of 
our paper is on the impact of feedback on the large-scale HI properties of 
galaxies, we summarize the modes of feedback that are implemented in 
{\texttt{GALFORM}} -- from SNe (\S~\ref{SNe}), AGN (\S~\ref{AGN}) and 
photo-ionization after cosmological reionization (\S~\ref{PH}).

\subsection{The {\texttt{GALFORM}} galaxy formation model}\label{GFM}

We use the version of {\texttt{GALFORM}} \citep[cf.][]{cole.etal.2000}
that is described in \citet{lagos.etal.2011a,lagos.etal.2011b} 
to predict the properties of galaxies forming and evolving in the
$\Lambda$CDM cosmology adopted for the Millennium Simulation 
\citep[cf.][]{springel.etal.2005}\footnote{Recall that the cosmological
parameters adopted for the Millennium Simulation are the total matter density
$\Omega_{\rm M}=0.25$, the baryon density $\Omega_{\rm b}=0.045$, 
the vacuum energy density $\Omega_{\Lambda}=0.75$, the Hubble parameter
$H_{0}=100h \,{\rm km s}^{-1}\,{\rm Mpc}^{-1}$ with $h$=0.73, the primordial 
scalar spectral index $n_{\rm s}=1$ and the fluctuation amplitude 
$\sigma_{8}=0.9$.}. 
{\texttt{GALFORM}} models the key physical 
processes of galaxy formation, including the gravitationally driven assembly 
of dark matter halos, radiative cooling of gas and its collapse to form 
centrifugally supported discs, star formation, and feedback from
supernovae (SNe) and active galactic nuclei (AGN).

\citet{lagos.etal.2011a} extended {\texttt{GALFORM}} by modelling the splitting
of cold gas in the interstellar medium (ISM) into its HI and H$_2$ components 
and by linking explicitly star formation to the amount of H$_2$ present in a 
galaxy. \citet{lagos.etal.2011b} compared the model predictions obtained using
empirically and theoretically derived star formation laws 
\citep[cf.][]{blitz.2006,krumholz.etal.2009} with a variety of observations 
(e.g. the HI mass function, $^{12}$CO (1-0) luminosity function, and 
correlations between the ratio H$_2$/HI and stellar and cold gas masses) and 
found that the empirical law of \citet{blitz.2006} (see also \citealt{leroy.etal.2008}) is 
favoured by these data. This law is of the form
\begin{equation}
\Sigma_{\rm SFR} = \nu_{\rm SF} \,\rm f_{\rm mol} \,\Sigma_{\rm gas},
\label{Eq.SFR}
\end{equation}
\noindent where $\Sigma_{\rm SFR}$ and $\Sigma_{\rm gas}$ are the surface
densities of the star formation rate (SFR) and total cold gas mass 
respectively, $\nu_{\rm SF}$ is the inverse of the SF timescale for the 
molecular gas and $\rm f_{\rm mol}=\Sigma_{\rm mol}/\Sigma_{\rm gas}$ is the
molecular to total gas mass surface density ratio. Importantly for the work
we present in this paper, \citet{lagos.etal.2011b} showed that the 
\citet{blitz.2006} law is able to reproduce the HI mass function at $z$=0 at 
intermediate and low HI masses. This is because it suppresses star formation 
in lower mass galaxies, thereby reducing SNe feedback and allowing these 
galaxies to retain larger gas reservoirs. Note that we use the 
\citet{lagos.etal.2011b} as the default model in this paper. 

\subsection{Supernovae (SNe) Feedback}
\label{SNe}

SNe act to reheat {cold gas} and eject it from the galaxy disc at a rate \begin{equation} 
\dot{M}_{\rm eject}=\beta\psi, \end{equation} where $\psi$ is the 
instantaneous star formation rate and $\beta$ regulates feedback efficiency,
 \begin{equation} \label{eq:beta} 
\beta=(V_{\rm disk}/V_{\rm hot})^{-\alpha_{\rm hot}}. \end{equation} \noindent 
Here $V_{\rm hot}$ and $\alpha_{\rm hot}$ are adjustable parameters 
that govern the strength of the SNe feedback \citep[cf.][]{cole.etal.2000} 
and $V_{\rm disk}$ is the circular velocity of the galactic disc; the values 
in the default \citet{lagos.etal.2011b} model are $V_{\rm hot}$=485 km/s and 
$\alpha_{\rm hot}$=3.2. We expect SNe feedback 
to be most effective in low-mass dark matter halos because gas is expelled 
efficiently from their shallow potential wells. From Eq~\ref{eq:beta}, the 
mass loading of the SNe wind is largest in such halos and this suppresses the 
formation of faint, low-mass galaxies and ensures that the faint end of the 
galaxy luminosity function is recovered by {\texttt{GALFORM}}, in good 
agreement with observational estimates 
\citep[e.g.][]{blanton.etal.2001,norberg.etal.2002}.

\subsection{AGN feedback}
\label{AGN}

SNe feedback is less effective in large circular velocity halos (according
to Eq~\ref{eq:beta}). The gas ejected from low mass halos can potentially
cool and turn into stars in more massive halos, generally leading to the 
formation of too many massive galaxies. In the absence of strong feedback in 
massive dark matter halos, {\texttt{GALFORM}} predicts an overabundance of 
bright galaxies and a galaxy luminosity function whose bright end slope is too 
steep \citep[cf.][]{benson.etal.2003a}. For this reason, 
\citet{bower.etal.2006} extended {\texttt{GALFORM}} to include AGN feedback, 
which suppresses cooling flows in massive halos. Physically this occurs 
because {AGN are fuelled by super-massive black 
holes that grow by luminous accretion}.

{In {\texttt{GALFORM}} AGN feedback is modelled by assuming that 
the hot gaseous atmospheres surrounding massive galaxies are in 
quasi-hydrostatic equilibrium when the cooling time at the cooling radius, 
$t_{\rm cool}(r_{\rm cool})$, exceeds a multiple of the free-fall time at the 
cooling radius, $t_{\rm ff}(r_{\rm cool})$, i.e.}

\begin{equation} 
  \label{eq:tcool}
  {
  t_{\rm cool}(r_{\rm cool})>{1 \over \alpha_{\rm cool}} t_{\rm ff}(r_{\rm cool}); 
  }
\end{equation} 
{ where $\alpha_{\rm cool}$ is a free parameter that regulates 
the strength of AGN feedback\footnote{{We note that Eq 2 of 
\citet{bower.etal.2006} contains a typographical error.}}; increasing 
$\alpha_{\rm cool}$ means that AGN feedback becomes effective in 
lower mass dark matter halos. We adopt a value of $\alpha_{\rm cool}$=0.58, 
which is the default in the \citet{lagos.etal.2011b} model. We define the 
cooling time at halocentric radius $r$ as}

\begin{equation}
  \label{taucool}
  {
  t_{\rm cool}(r)={3 \over 2}{ \bar{\mu}m_{p}k_{B}T_{\rm gas} \over \rho_{\rm gas}(r)\Lambda(T_{\rm gas},Z_{\rm gas})},}
\end{equation}
{where $\rho_{\rm gas}$, $T_{\rm gas}$ and $Z_{\rm gas}$ are the 
gas density, temperature and metallicity respectively, $\Lambda$ is the cooling 
function\footnote{We use the tabulated cooling functions of 
\citet{sutherland.dopita.1993}.}, $\bar{\mu}$ is the mean molecular weight, 
$m_p$ the proton mass and $\rm k_{\rm B}$ is Boltzmann's constant. We define 
the free-fall time at halocentric radius $r$ by}
\begin{equation}\label{taufreefall}
  {
  t_{\rm ff}(r)=\int^{r}_{0}\left[ \int^{y}_{r}-{2GM(x) \over x^{2}} {\rm d}x \right]^{-1/2}{\rm d}y.
  }
\end{equation}
{where $M(r)$ is the total mass (baryonic and dark matter) 
interior to $r$. The cooling radius, 
$r_{\rm cool}$, is the radius at which $t_{\rm cool}$ matches the age of the halo,
which is obtained from the halo merger history.}

{We note that the model predicts that generally 
black holes gain most of their mass by 
accretion of gas driven to the centre of the galaxy via disc instabilities 
\citep[cf.][]{fanidakis.etal.2011}. Accretion of gas that cools from the hot 
atmosphere is relatively unimportant until late times, when the growth rate of 
the black hole scales as $L_{\rm cool}/c^2$, where
$L_{\rm cool}$ is the luminosity of the cooling gas and $c$ is the speed of light.}

\subsection{Photo-ionization feedback}
\label{PH}

Photo-ionization is predicted to have a dramatic impact on star formation in 
low-mass galaxies. This is because the presence of a photo-ionizing background 
both modifies the net cooling rate of gas in halos by removing the ``hydrogen peak''
in the cooling curve \citep[cf. Fig. 1 of][]{benson.etal.2002} and increases
the temperature of the intergalactic medium (IGM) such that its thermal pressure 
prevents gravitational collapse of baryons onto low-mass halos 
\citep[e.g.][]{efstathiou.1992,okamoto.etal.2008}. As a result, only those 
low-mass halos that contained cold gas prior to re-ionization can form stars 
\citep[e.g.][]{hoeft.etal.2006}. {\texttt{GALFORM}} includes the 
\citet{benson.etal.2002} prescription for suppressing the cooling of 
halo gas onto the galaxy -- this occurs if the host halo's circular 
velocity $V_{\rm circ}$ lies below a threshold $V_{\rm cut}$ at redshift 
$z < z_{\rm cut}$. The values in the default \citet{lagos.etal.2011b} model 
are $V_{\rm cut}$=30 km/s and {\bf $z_{\rm cut}=10$}. The default $V_{\rm cut}$ is in good agreement 
with the results of hydrodynamical simulations by \citet{hoeft.etal.2006} and 
\citet{okamoto.etal.2008}. 

\section{The Galaxy Luminosity Function and HI Mass Function}
\label{sec:himf}

In this section we investigate how the galaxy luminosity function in 
$b_{\rm J}$ band and the HI mass function at $z$=0 are influenced by 
feedback from supernovae (SNe), active galactic nuclei (AGN) and 
a photo-ionizing background. We use Monte Carlo halo merger trees generated 
using the algorithm described in \citet{parkinson.etal.2008} to predict galaxy 
properties. The Monte Carlo trees are tuned to match $N$-body merger trees 
extracted from the Millennium Simulation. Note that we express HI masses in 
units of $h^{-2} \rm M_{\odot}$ rather than $h^{-1} \rm M_{\odot}$, the mass unit 
used in simulations. This ensures that the observational units (which depend 
upon the square of the luminosity distance) are matched, but it introduces an 
explicit dependence on the dimensionless Hubble parameter, for which we assume 
the value of $h$=0.73 used in the Millennium simulation. We refer the reader 
to \S~2.1 of \citet{lagos.etal.2011b} for further discussion of this approach.\\

Intuitively, we expect SNe feedback to be particularly damaging in low-mass 
galaxies because gas is expelled easily from their shallow potential wells 
and so star formation is shut down. In contrast, SNe feedback is ineffective in 
high-mass galaxies, which reside at the centres of massive dark matter 
halos with deep potential wells; here a far more energetic source of 
feedback, namely AGN heating, is required, which acts to suppress 
the formation of very massive bright galaxies. The photo-ionizing 
background suppresses the collapse of gas onto low mass dark 
matter halos, thereby preventing galaxies from forming. The influence of
these distinct processes on the galaxy luminosity function is straightforward
to predict -- photo-ionization and SNe feedback will govern the amplitude and 
shape of the faint end, while AGN heating dictates the amplitude and shape of 
the bright end. Predicting their impact on the HI mass function is less 
straightforward -- as noted in \citet{lagos.etal.2011b}, the most HI-rich 
galaxies are likely to be low-mass systems, in which case the imprint of
SNe feedback may span the mass function, whereas the influence of AGN heating 
may be negligible.

\begin{figure*}
  \includegraphics[width=7.25cm]{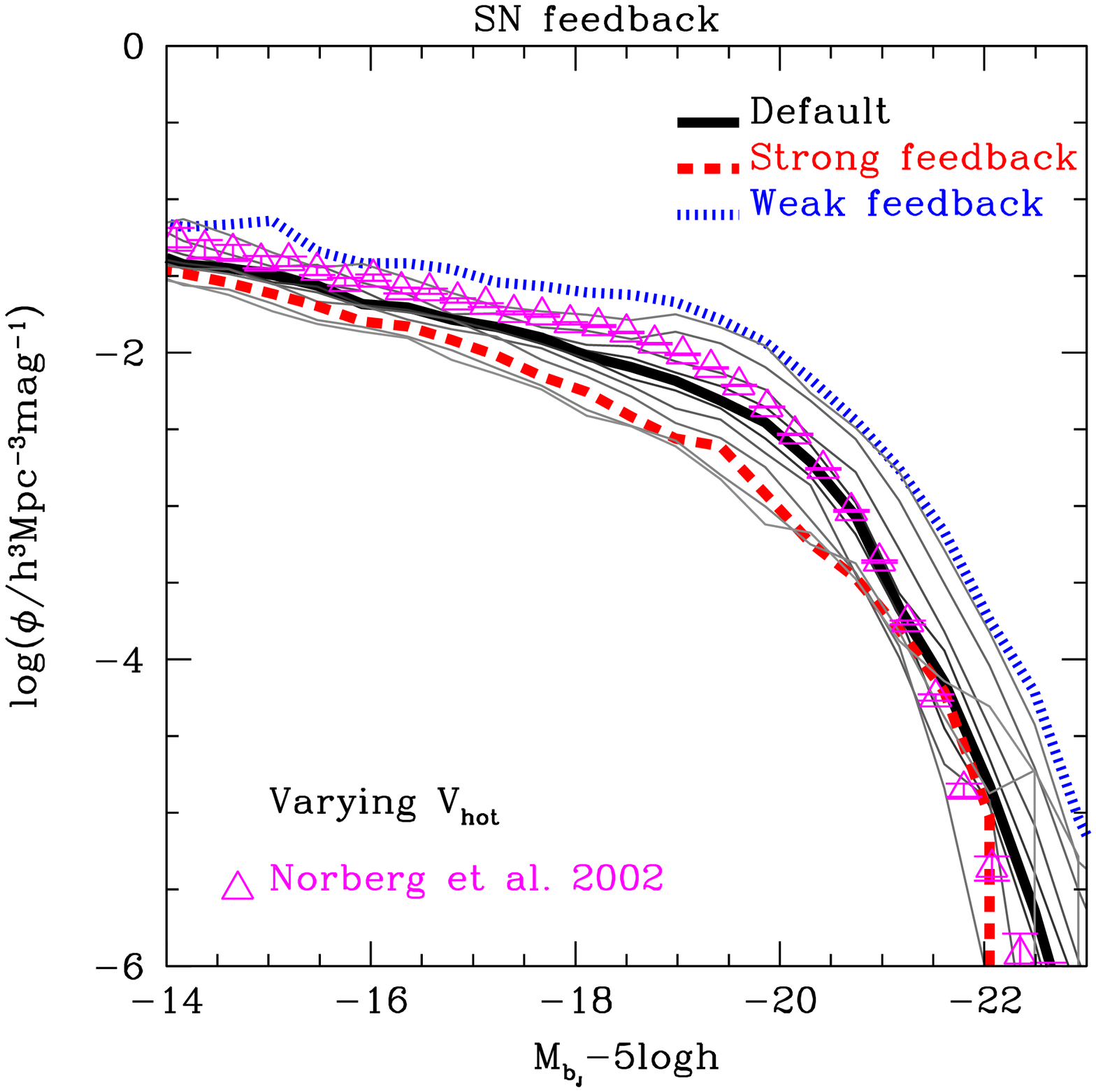}
  \includegraphics[width=7.25cm]{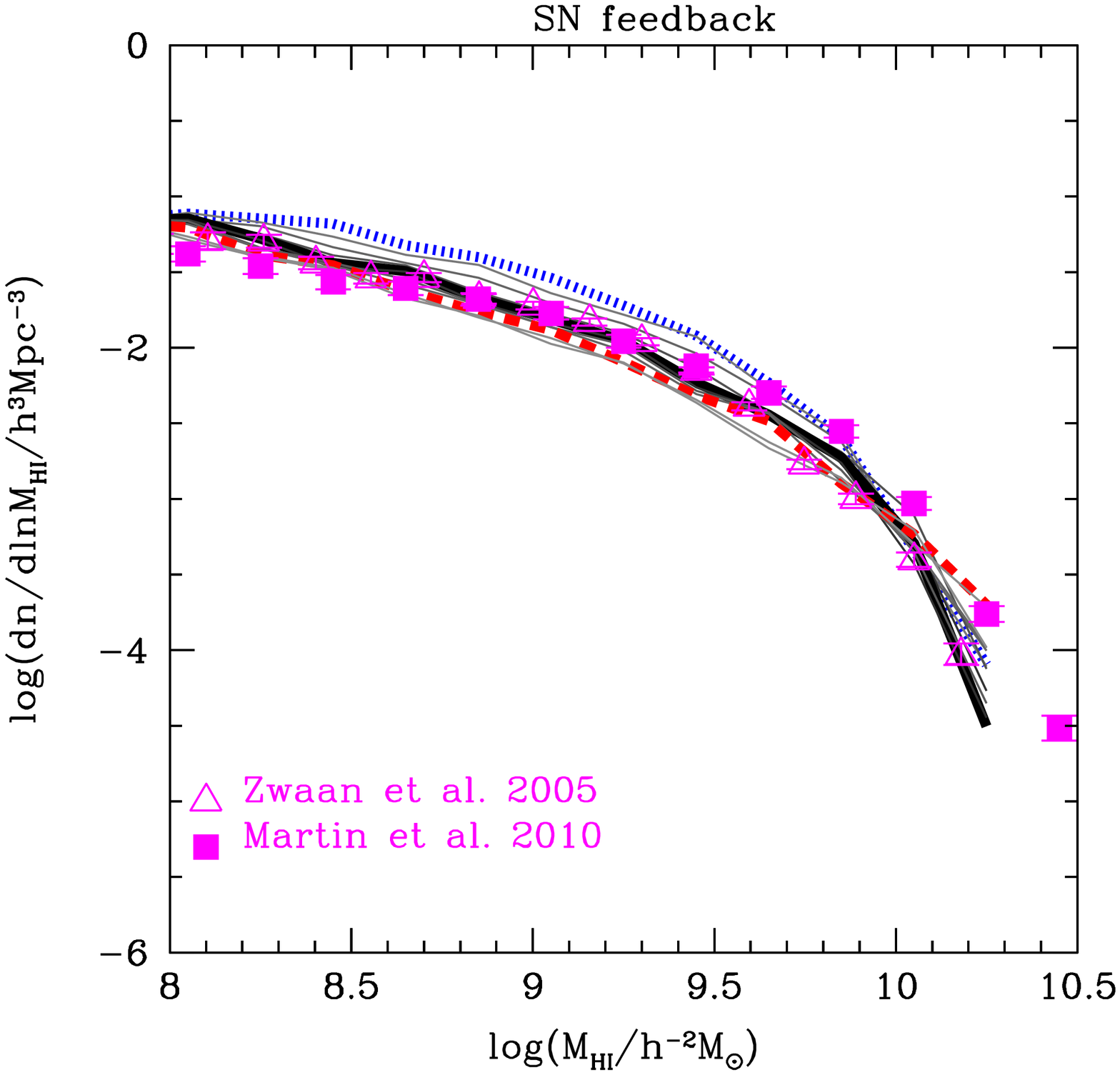}
  \includegraphics[width=7.25cm]{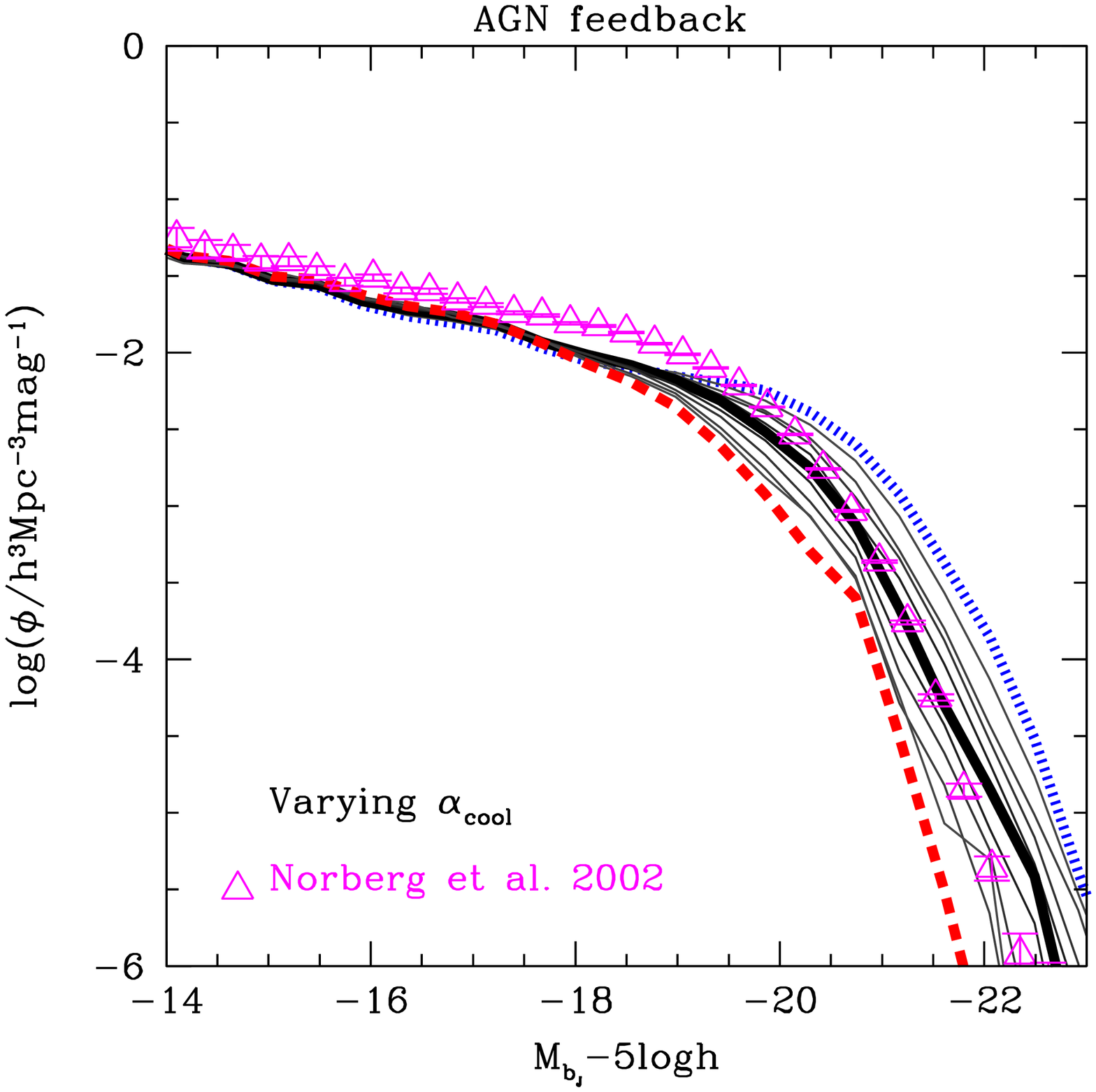}
  \includegraphics[width=7.25cm]{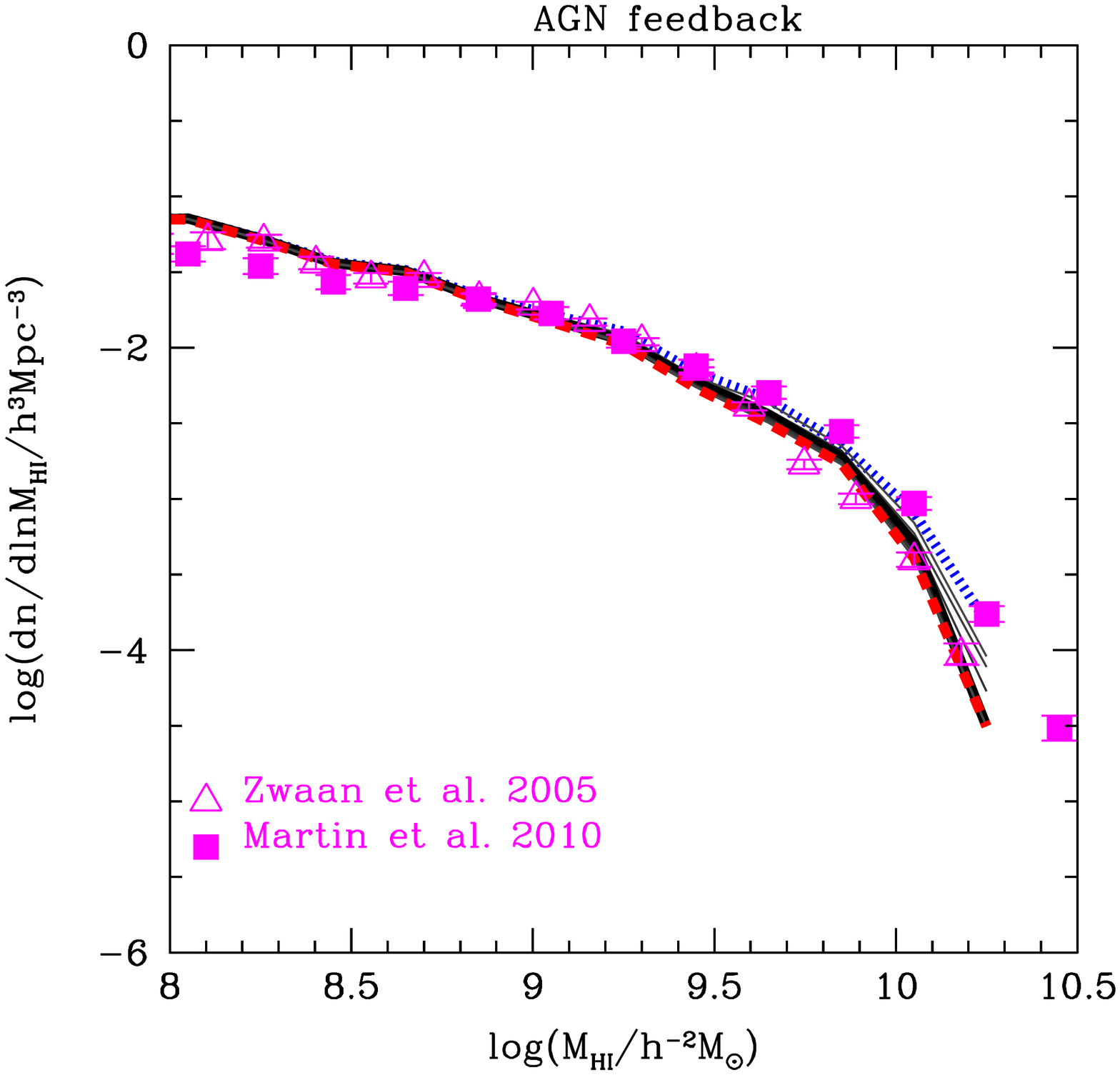}
  \includegraphics[width=7.25cm]{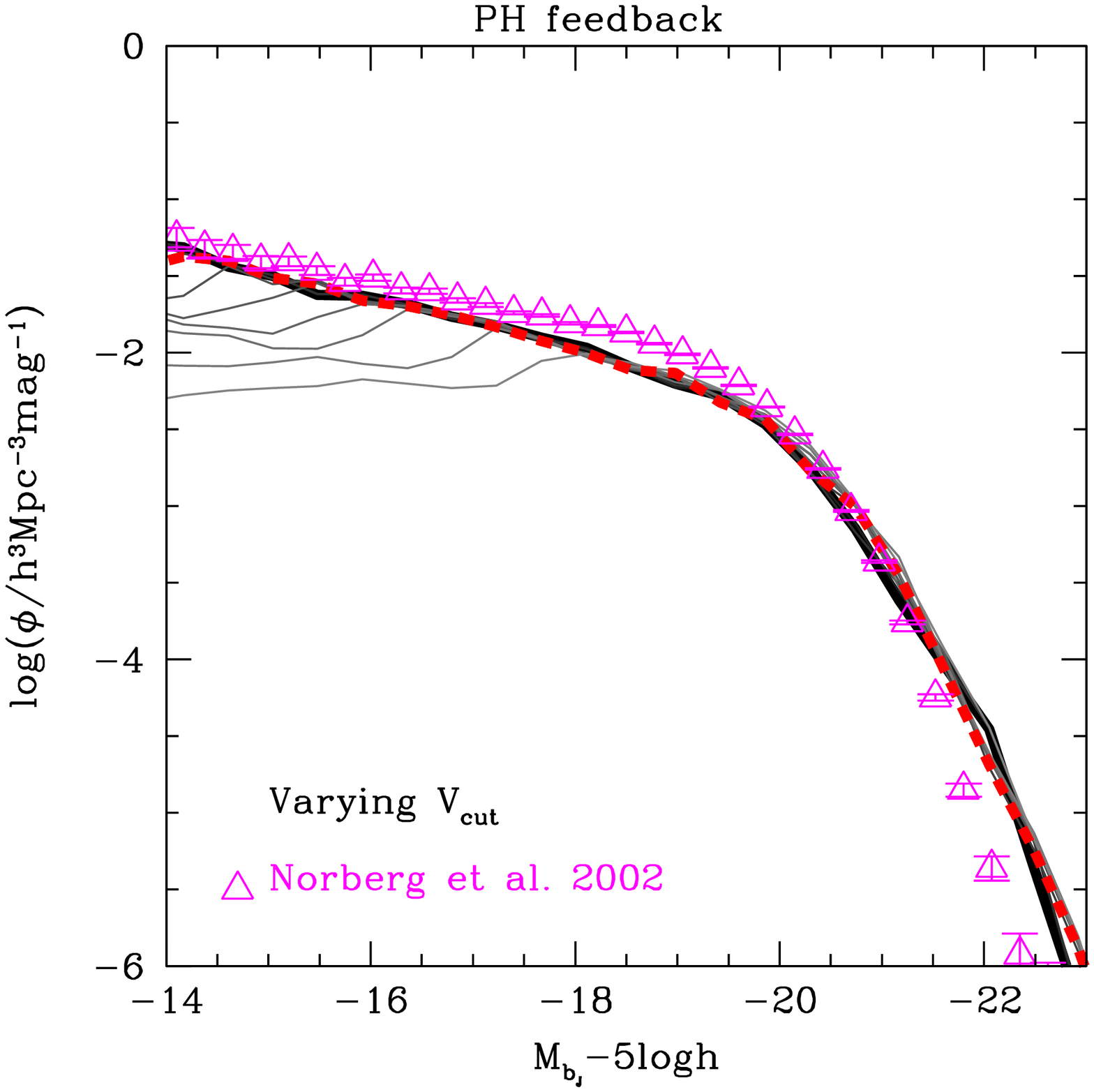}
  \includegraphics[width=7.25cm]{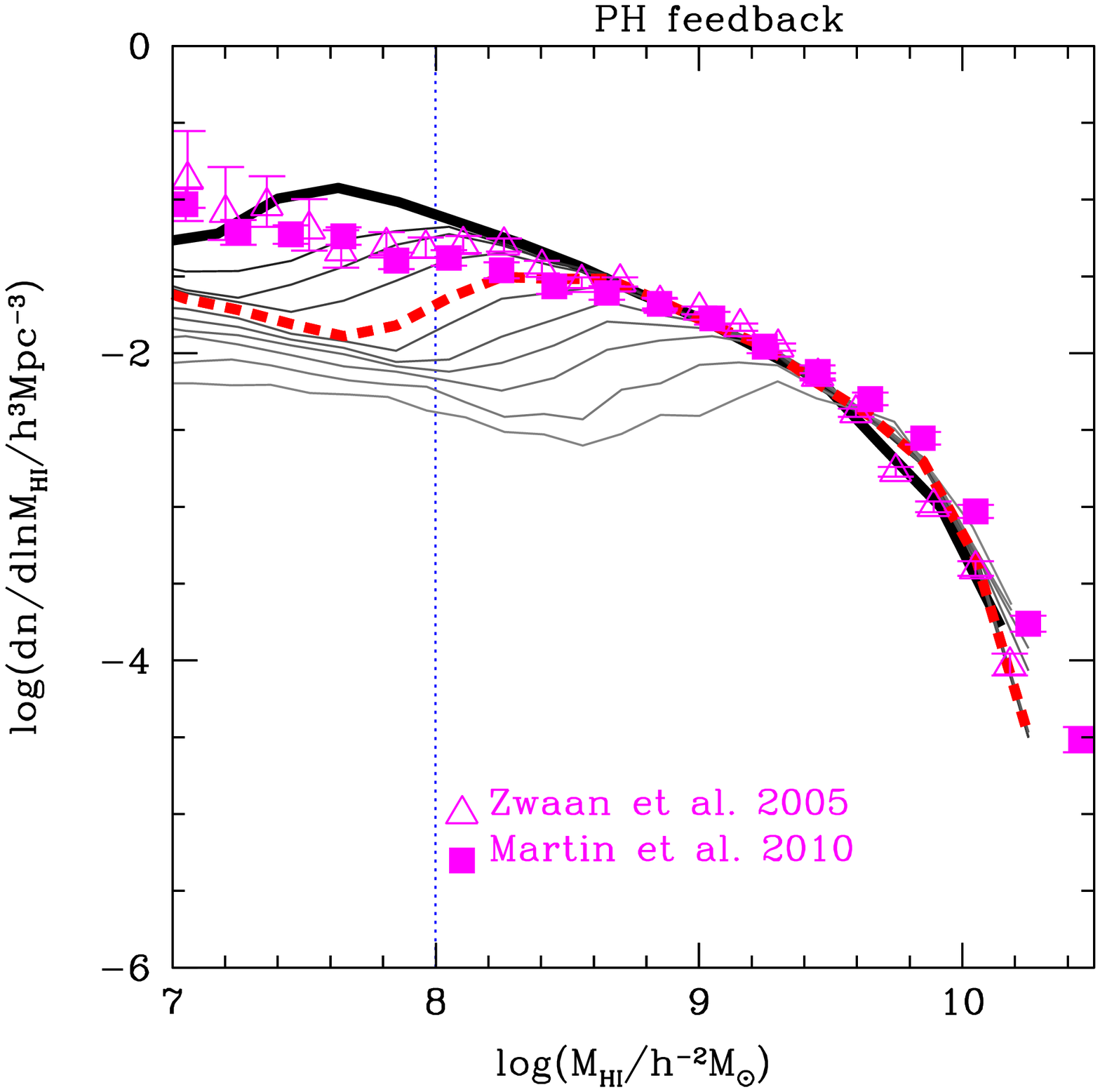}
  \vspace{-0.5cm}
  \caption{Impact of feedback from SNe, AGN and a photo-ionizing background 
    (top, middle and bottom rows) on the predicted $b_{J}$-band galaxy 
    luminosity function (left panels) and HI mass function (right panels); 
    symbols correspond to data from the 2dF Galaxy Redshift Survey 
    \citep[open triangles; { cf. Figure 11 and Table 1 of} ][]{norberg.etal.2002}, 
    HIPASS \citep[open triangles; cf.][]{zwaan.etal.2005} and ALFALFA
    \citep[filled squares; cf.][]{martin.etal.2010}. 
    SNe strength varies between $300 {\rm km/s} \leq V_{\rm hot} 
    \leq 700 {\rm km/s}$, assuming fixed $\alpha_{\rm hot}$;
    AGN strength between $0.4 \leq \alpha_{\rm cool} \leq 0.8$; and 
    photo-ionization strength between 
    $30 {\rm km/s} \leq V_{\rm cut} \leq 90 {\rm km/s}$. The darker the 
    line, the closer the strength of the feedback parameter to its 
    default value. Red dashed and blue dotted curves correspond to models with strong
    and weak feedback; the dotted vertical line in the lower right panel indicates the lower HI mass limit in the galaxy formation model we use them as reference models in 
    \S~\ref{sec:clustering}.}
  \label{LFDCOS}
\end{figure*}

We investigate these ideas in Fig \ref{LFDCOS}, in which we explore the 
impact of SNe driven winds (upper panels), AGN heating (middle panels) and 
a photo-ionizing background (lower panels) on the galaxy luminosity function in $b_{\rm J}$ 
band (left panels) and HI mass function (right panels) at $z$=0. For
comparison we plot the $b_{J}$ band galaxy luminosity function (open
triangles) estimated from the 2 degree Field Galaxy Redshift Survey (2dFGRS) 
by \citet{norberg.etal.2002} in the left panels, and the HI mass functions
deduced from the HIPASS (open triangles) and ALFALFA (filled squares)
surveys by \citet{zwaan.etal.2005} and \citet{martin.etal.2010} respectively.

\paragraph*{Impact of SNe feedback:} We vary the parameter $V_{\rm hot}$, 
which governs the mass ejection rate, $\dot{M}_{\rm eject}$, assuming a fixed
value of $\alpha_{\rm hot}$, between $300 {\rm km/s}$ (less effective) and 
$700 {\rm km/s}$ (more effective); recall its default value is 485 km/s.
{By varying $V_{\rm hot}$ we are in essence changing the mass of 
the system in which SNe feedback is effective. On the other hand varying $\alpha_{\rm hot}$ 
would result in the same galaxies being affected by SNe, but with different mass 
loading in the winds of reheated material.} 
The upper panels indicate that varying $V_{\rm hot}$ in this way leads 
to systematic and measurable offsets in both the luminosity function and 
HI mass function -- stronger (weaker) SNe feedback decreases (increases) the 
number of galaxies of a given luminosity (by $\sim$ 0.5-1 dex) or HI mass 
(by $\sim$ 0.5 dex). Interestingly, weaker feedback leaves the shape of the
luminosity and HI mass functions broadly unaffected, whereas stronger feedback 
leads to slightly steeper slopes at the faint/low-mass and bright/high mass 
ends.
Red dashed and blue dotted curves correspond to models with 
{strong ($600 {\rm km/s})$} and 
{weak ($300 {\rm km/s})$} feedback; 
we use these as reference models in \S~\ref{sec:clustering}, where we examine 
the clustering of HI-rich and HI-poor galaxies.

\paragraph*{Impact of AGN feedback:} Here we vary the parameter
$\alpha_{\rm cool}$ between $0.4 \leq \alpha_{\rm cool} \leq 0.8$; recall its 
fiducial value is $\alpha_{\rm cool}=0.58$. Increasing $\alpha_{\rm cool}$ 
decreases the halo mass at which AGN heating starts to become effective. The 
middle panels of Fig \ref{LFDCOS} demonstrate that AGN heating has little 
influence beyond the bright end of the galaxy luminosity function or the 
high-mass end of the HI mass function, regulating both the sharpness of the 
transition between faint/bright and low/high mass galaxies and the position at 
which it occurs. The number of bright galaxies is very sensitive to the 
strength of AGN feedback, varying by as much as a factor of $\sim$ 
{100} for the most luminous systems. Interestingly AGN feedback 
also affects the numbers of galaxies with high HI masses, but the effect is 
only apparent in the cases of weakest feedback. {Red dashed and blue dotted
curves correspond to models with strong ($\alpha_{\rm cool} = 0.8$) 
and weak ($\alpha_{\rm cool} = 0.4$)} AGN feedback, 
which, as before, we use as reference models in \S~\ref{sec:clustering}.

\paragraph*{Impact of Photo-ionization:} Two parameters, $V_{\rm cut}$ and 
$z_{\rm cut}$, govern photo-ionization in the model. We find that $z_{\rm cut}$, 
the redshift at which reionization occurs, has little influence on either the 
galaxy luminosity function or HI mass function at $z$=0 -- 
{varying $z_{\rm cut} $ between
$6 \leq z_{\rm cut} \leq 12$, the range suggested by measurements of polarisation 
in the cosmic microwave background radiation and the Gunn-Peterson effect in distant 
quasars \citep[e.g.][]{greiner.etal.2009,komatsu.etal.2011}, produce luminosity and 
HI mass functions that are indistinguishable, to within the line width. For this reason we 
concentrate on $V_{\rm cut}$, which we vary between 
$30 {\rm km/s} \leq V_{\rm cut} \leq 90 {\rm km/s}$. The minimum progenitor mass we consider is 
$5\times 10^{8} h^{-1} \odot$ to explore low $V_{\rm cut}$ values (see also HI mass function in figure.~\ref{LFDCOS} which
has lower HI mass limit than other cases of HI mass function, indicated by the vertical dotted line).} Increasing $V_{\rm cut}$ 
suppresses cooling in all halos with circular velocities 
$V_{\rm cir} \leq V_{\rm cut}$ at $z < z_{\rm cut}$ and its impact is most 
pronounced at the faint end of the galaxy luminosity function and the low-mass 
end of the HI mass function. {The red dashed curve corresponds to a 
model with strong (50${\rm km/s}$) photoionization feedback, which we use as a reference model 
in \S~\ref{sec:clustering}.}

\paragraph*{Constraints on Galaxy Formation Models:} What information can we 
glean from the galaxy luminosity and HI mass functions about the various modes 
of feedback? {We investigate this using the galaxy luminosity 
function of \citet{norberg.etal.2002} and the HI mass function of
\citet{zwaan.etal.2005}. Although the \citet{martin.etal.2010} ALFALFA mass 
function extends to lower HI masses, it has more limited sky coverage than the 
\citet{zwaan.etal.2005} HIPASS mass function and so is more susceptible to the 
effects of cosmic variance.} We quantify the constraints offered by the 
observational data by calculating the ratio of likelihoods $L/L_{\rm default}$ for 
each models with respect to default model,

\begin{equation}
  \frac{L}{L}_{\rm default}=\frac{\Sigma | \log(\mathrm{od_{i}})-\log(\mathrm{md_{i}}) |}{\Sigma | \log(\mathrm{od_{i}})-\log(\mathrm{md_{i}})|}_{\rm default}.
  \label{Chieq}
\end{equation}

\begin{figure*}
\includegraphics[width=5.5cm]{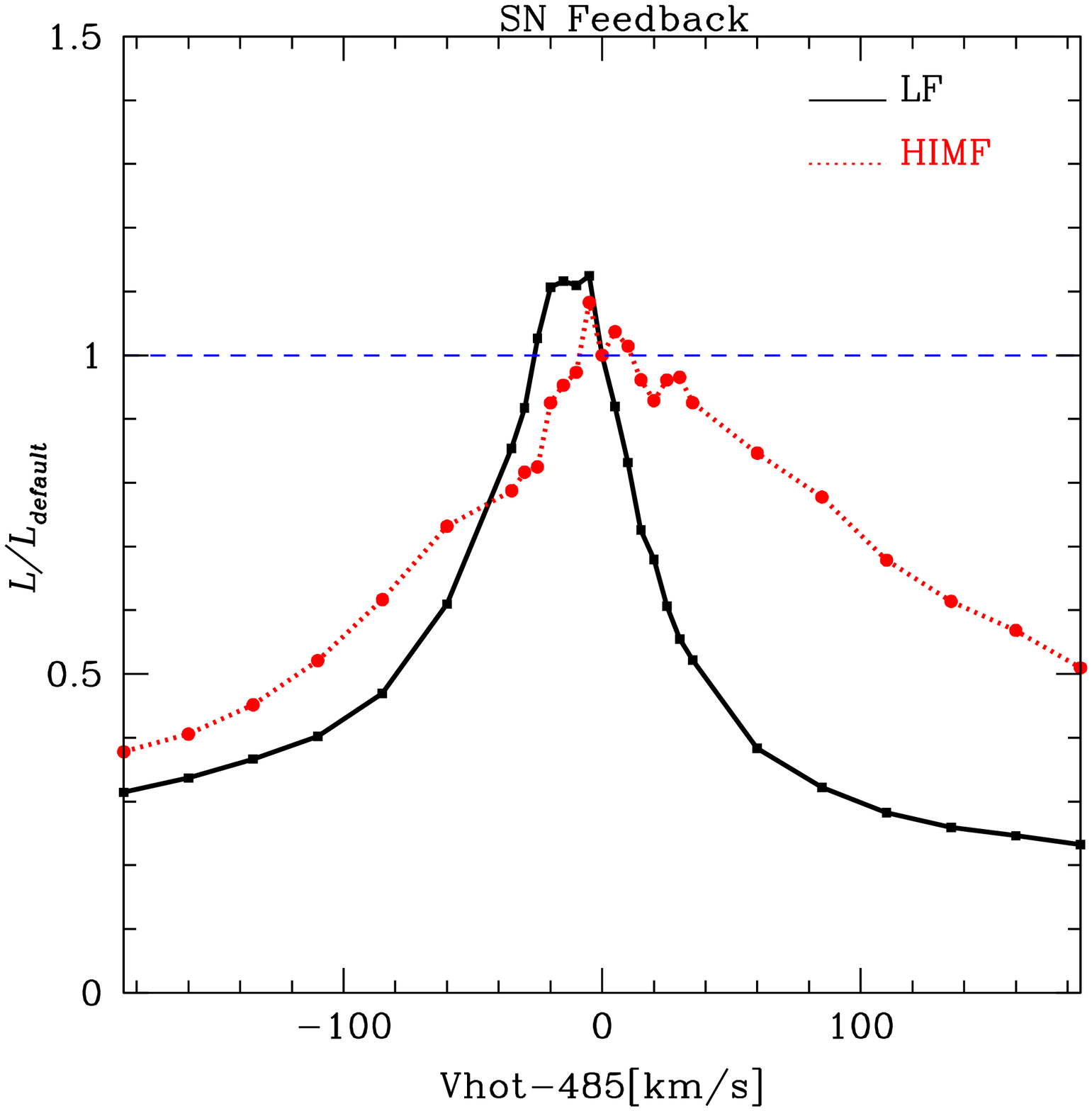}
\includegraphics[width=5.5cm]{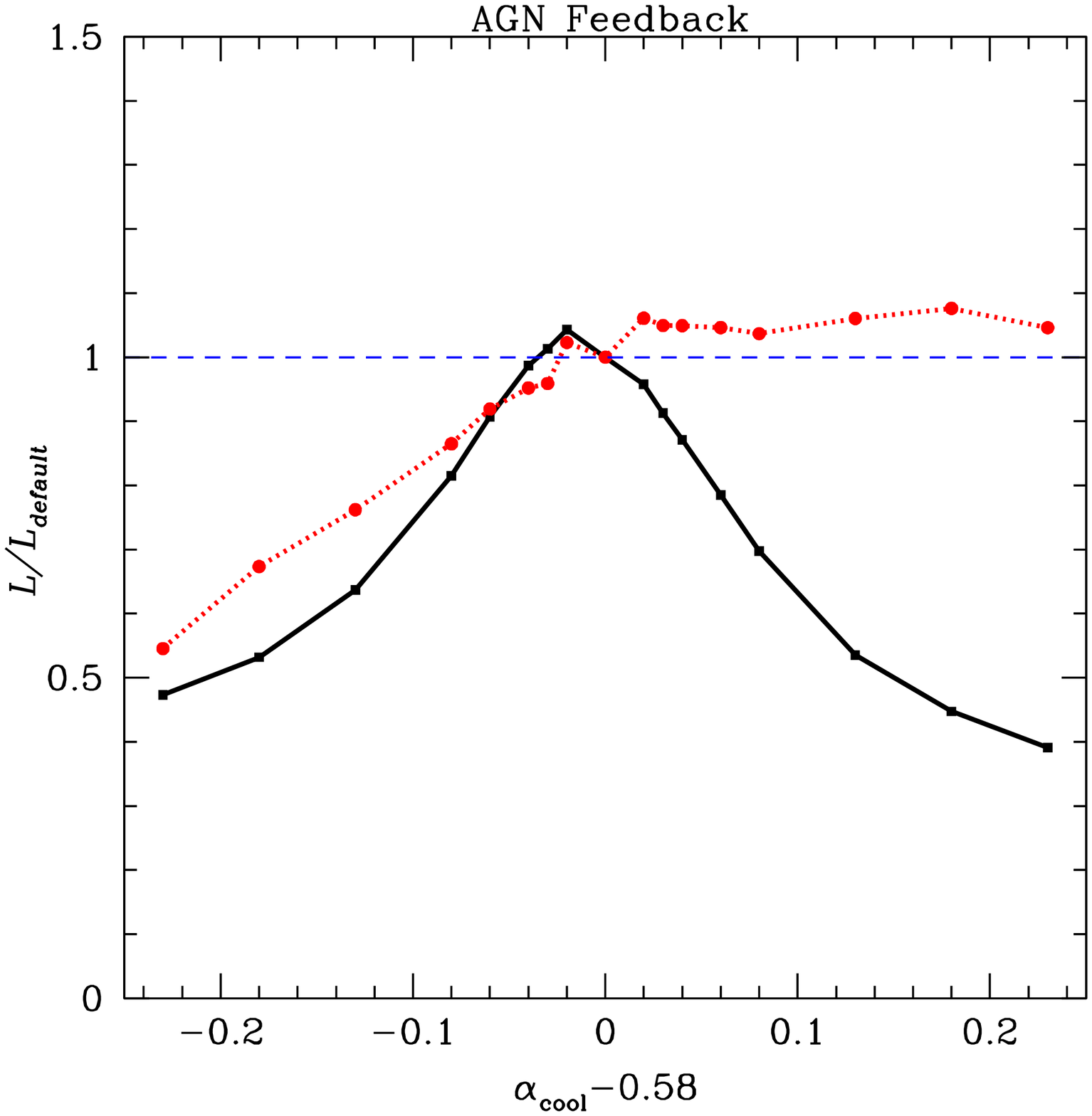}
\includegraphics[width=5.5cm]{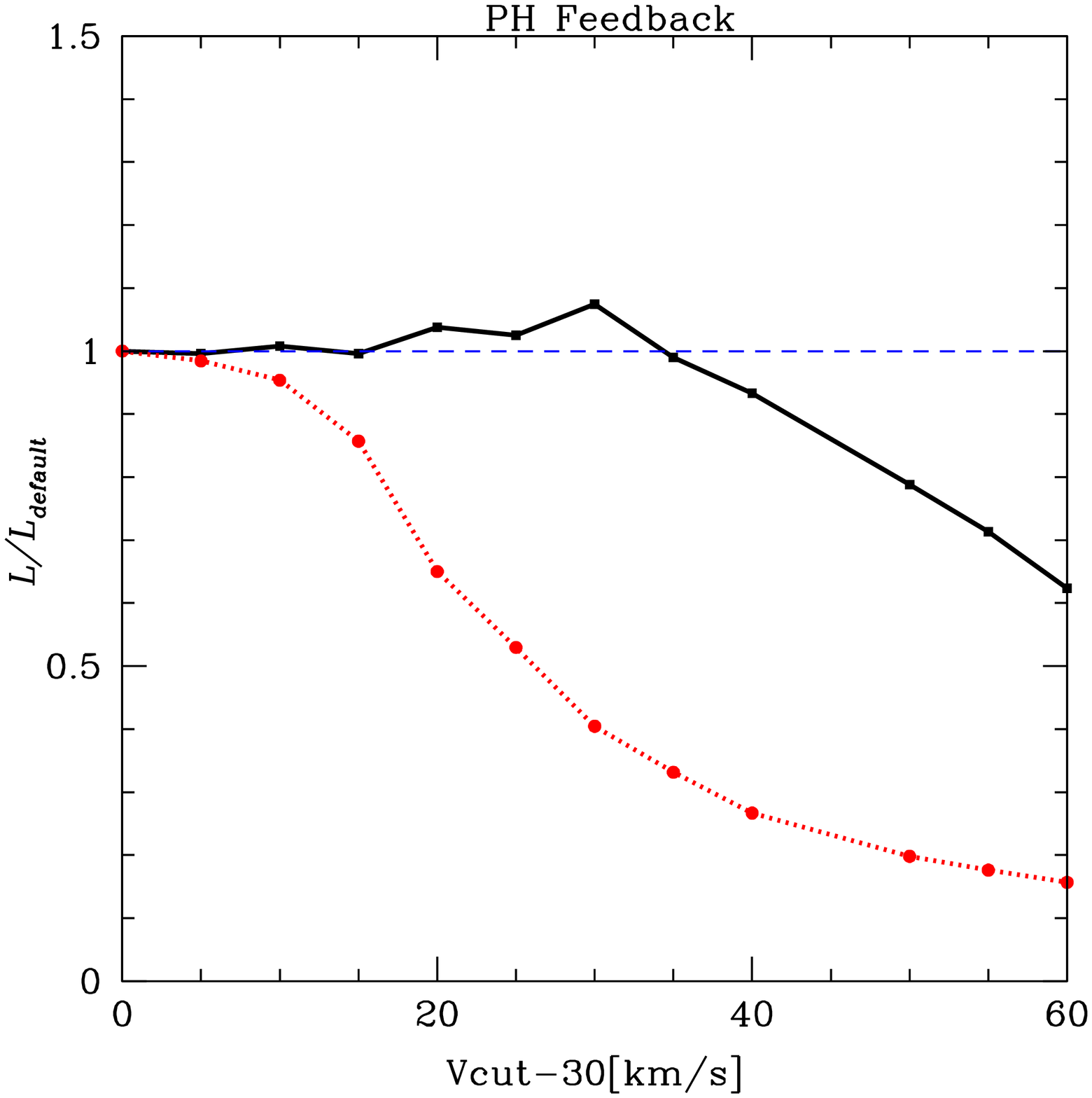}
\caption{{Values of $L/L_{\rm default}$ obtained for 
    the SNe, AGN and photo-ionization feedback models (left, middle and right 
    panels), where $L/L_{\rm default}$ is calculated using equation~\ref{Chieq}. 
    Solid (dotted) curves show how $L/L_{\rm default}$ for the galaxy luminosity 
    function in $b_{\rm J}$ band (HI mass function) as we vary systematically
    the SNe, AGN and photo-ionization strength; filled squares (circles) 
    indicate the values for the models that have been run. For reference we 
    show also a dashed horizontal lines corresponding to 
    $L/L_{\rm default}$=1. See text for further details.}}
\label{CHI}
\end{figure*}

\noindent {Here we sum over the difference in $\log$ values of the 
observational data (${\bf od_{i}}$) and the predictions (${\bf md_{i}}$) for each
model and compare with the sum over this difference for the default model. Our 
results are plotted in Fig~\ref{CHI}, where solid (dotted) curves indicate how 
$L/L_{\rm default}$ varies with SNe, AGN and photo-ionization feedback strength (left, middle and right
panels) for the galaxy luminosity function (HI mass function).}

{Formally, the models give a poor fit to the observational data. There are 
several reasons for this. Firstly, the parameters we vary are physical, not 
statistical, and affect the predictions of the model in a complex way, due to 
the interplay between different physical processes. With a parametric
functional form, such as the Schechter function, the consequences of changing 
a parameter are transparent. Secondly, GALFORM predicts a whole range of galaxy
properties, and the luminosity function or HI mass function represent just one 
way of describing the population of model galaxies. The parameters of the 
default model were chosen as a compromise in reproducing a variety of 
observational data, not just the luminosity function or mass function. If 
we chose to fit the data considered here in isolation, with complete disregard
for any other dataset, it would be possible to improve the fits. Nevertheless, 
it is not clear that we would obtain a formally acceptable fit. Again, the 
reason for this is that GALFORM is a physical model rather than a parametric 
fitting formula. The shortcomings of the model could be encoded in a 
''model discrepancy" that broadens the errors in a $\chi^2$ fit (see 
\citealt{bower.etal.2010} for a more formal approach to this). Finally, the 
observational errors do not include all of the contributions (e.g. sample 
variance) and exclude systematics (e.g. the k-correction in the case of the 
luminosity function). The key point we aim to get across from this plot is 
the shape of the $L/L_{\rm default}$ curves: How rapidly do we move 
away from the quality of the fit in the default model on varying the parameter?  
If the value of $L/L_{\rm default}$ varies rapidly on comparing the
model predictions to a particular dataset, then that dataset could in 
principle provide a tighter constraint on the parameter value.}

We find that the galaxy luminosity function shows a strong dependence on the 
strength of SNe feedback and a 
moderate dependence on the strength of AGN feedback, with a narrow range of 
preferred values close to their default values; the HI mass function offers
a weak constraint on SNe feedback and no constraint on AGN feedback. In 
contrast the HI mass function 
is sensitive to the strength of the photo-ionizing background, and indicates 
that values of $V_{\rm cut}$ larger than $\sim 70\,\rm km/s$ can be ruled out. 

Smaller values of $V_{\rm cut}$ appear to have the same likelihood, but this 
can be understood as an artifact of the effective resolution of our merger 
trees. We use Monte Carlo merger trees to explore the impact of varying 
$V_{\rm cut}$, as this allows us to vary the mass resolution of the progenitor 
halos to ensure that we can resolve the halos which would be affected by the 
choice of this parameter. The minimum progenitor mass we consider is 
$5\times 10^{8} h^{-1} \odot$. We can obtain tighter constraints on the minimum 
value of $V_{\rm cut}$ by using
Monte Carlo trees with higher effective resolution, but we defer this to a 
forthcoming paper. Our results suggest that the HI mass function, measured at 
$z$=0, can provide an important constraint on the minimum halo mass that 
contributed to reionization, and complements estimates estimates of the 
minimum halo mass based on high 
redshift observations 
\citep[e.g.][]{choudhury.etal.2008,srbinovsky.wyithe.2010,munoz.loeb.2011}

\section{Spatial Clustering of HI in Galaxies}
\label{sec:clustering}

We now examine how feedback influences the spatial distribution of HI 
galaxies, which we quantify using the 2-point correlation function and 
the halo occupation distribution \citep[hereafter HOD; 
cf.][]{benson.etal.2000,peacock.smith.2000,berlind.weinberg.2002}.
{The clustering predictions for $b_{\rm J}$-band selected 
samples were studied in Kim et~al. (2009).}
The HOD gives the mean number of galaxies as a function of halo mass, 
assuming pre-defined selection criteria (e.g. colour, cold gas mass), 
and can be separated into contributions from central galaxies and their 
satellites. It is an output from the semi-analytical model. HODs for 
optically selected samples of galaxies have been studied extensively 
and are typically parameterized as step functions, which represent 
central galaxies, and power-law components, which represent satellites 
\citep[e.g.][]{peacock.smith.2000,berlind.weinberg.2002}. In contrast, 
relatively few studies have been made of HODs for HI selected samples of 
galaxies \citep[e.g.][]{wyithe.etal.2010,marin.etal.2010,kim.etal.2011} but 
they indicate that parameterizations of the kind adopted for optically 
selected galaxy samples need to be modified for HI selected samples 
\citep[cf.][]{kim.etal.2011}.
 
Note that we use $N$-body halo merger trees drawn from the Millennium 
Simulation to obtain the spatial information required to study clustering.
The resolution limit of the Millennium Simulation imposes a limit on the halo
mass that we can reliably resolve of $10^{10} h^{-1} \rm M_{\odot}$, which in
turn sets a limit of $M_{\rm HI} \sim10^{8.5}h^{-2}\rm M_{\odot}$ on the HI mass
that we can track.\\ 
 
\begin{figure*}
\includegraphics[width=7.25cm]{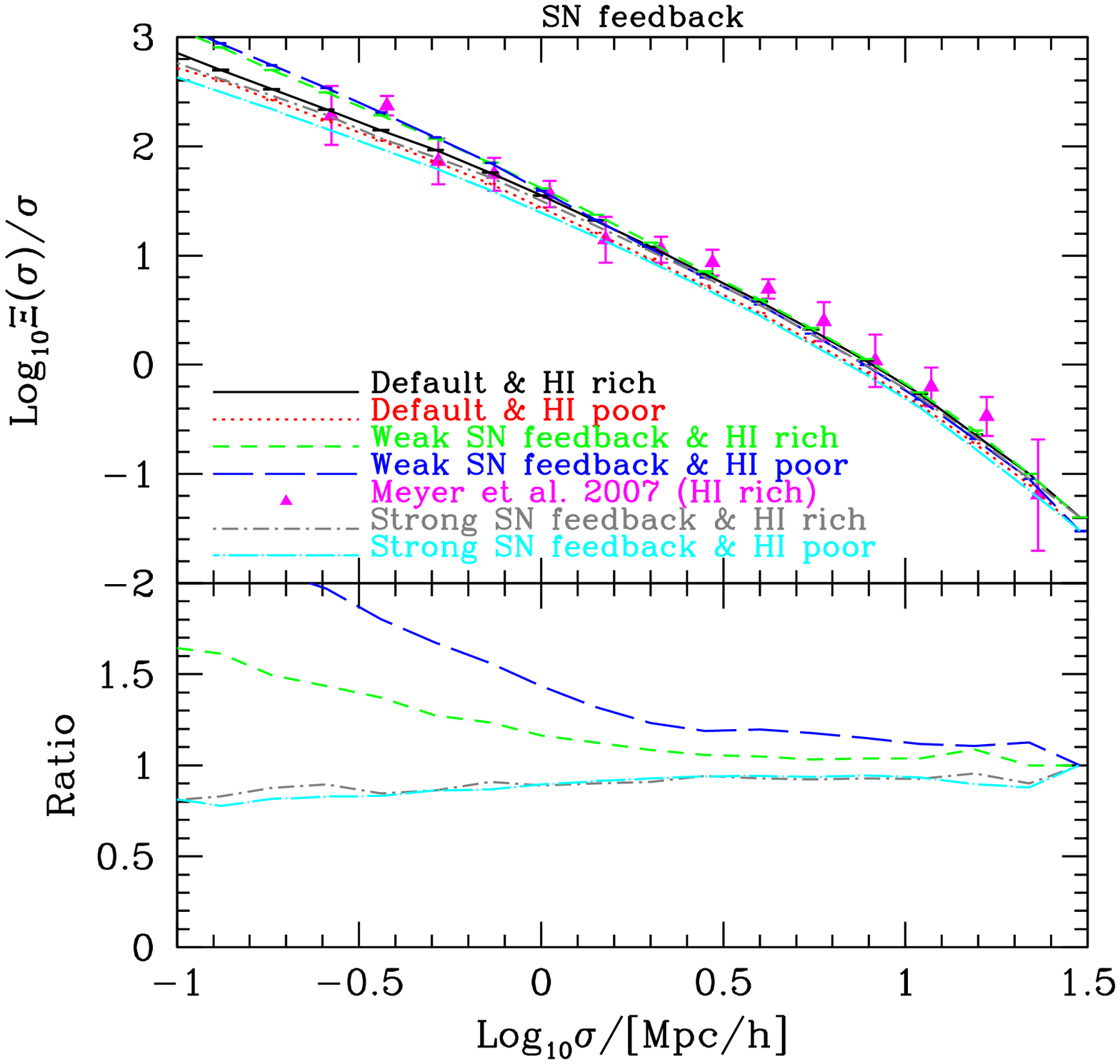}
\includegraphics[width=7.25cm]{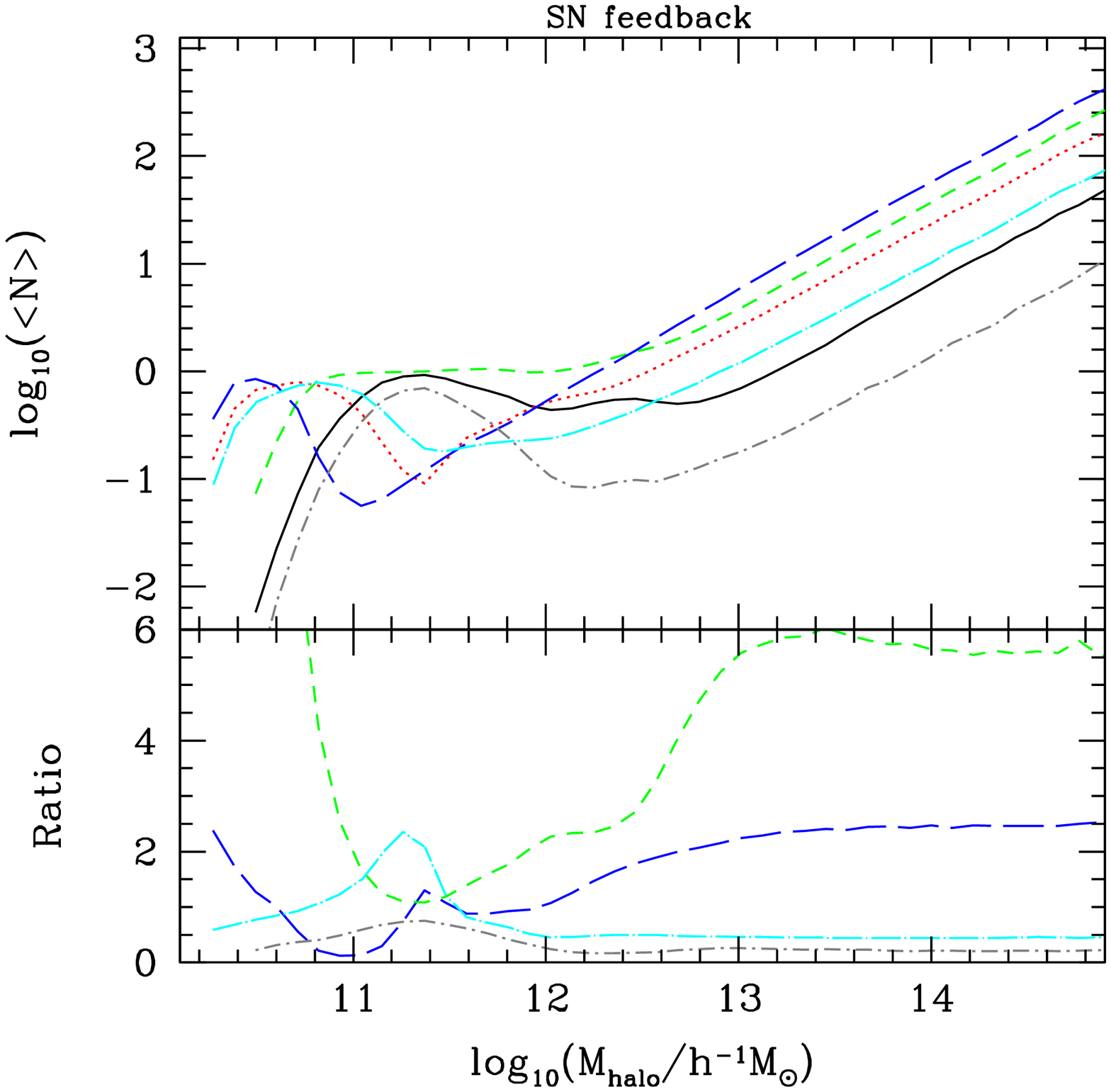}
\includegraphics[width=7.25cm]{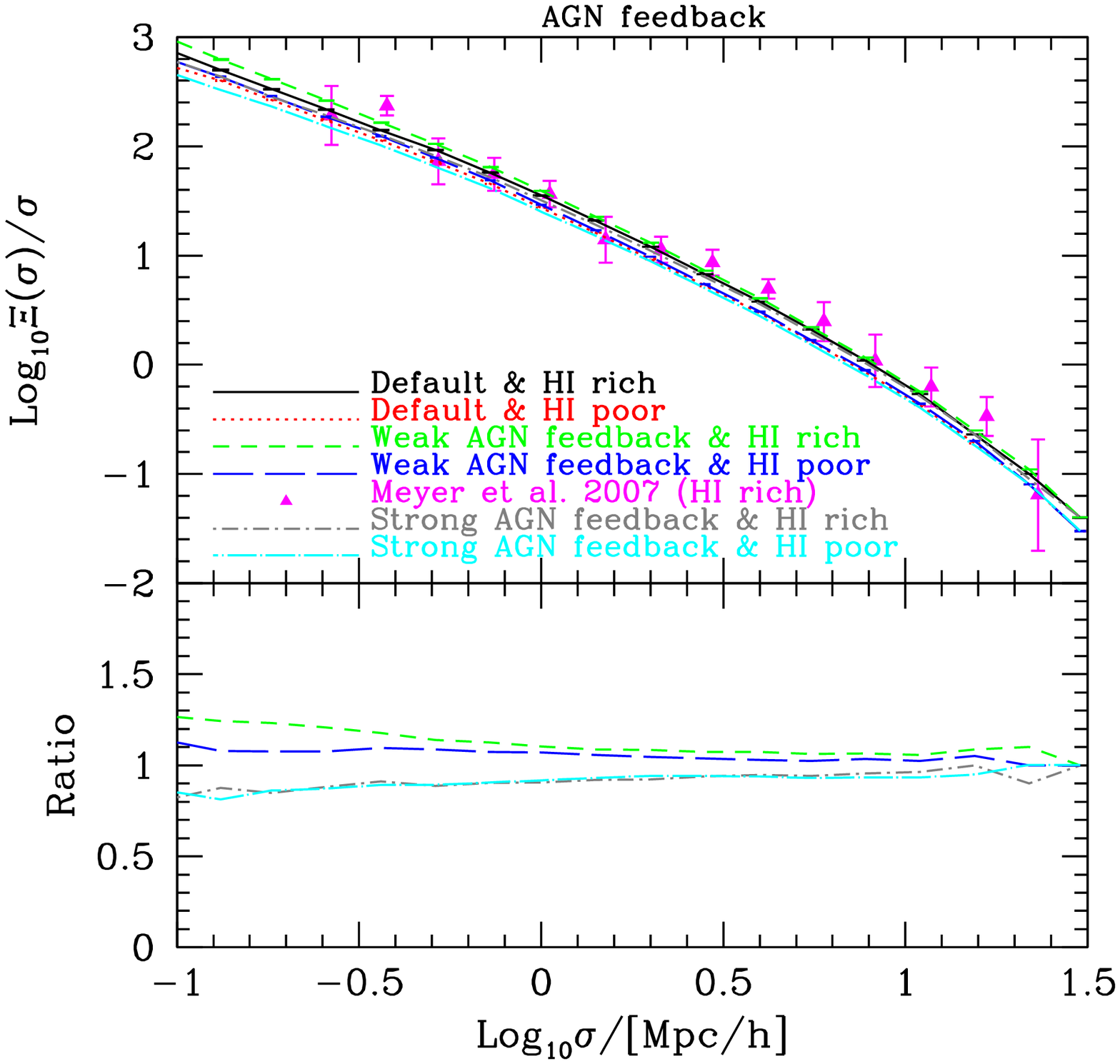}
\includegraphics[width=7.25cm]{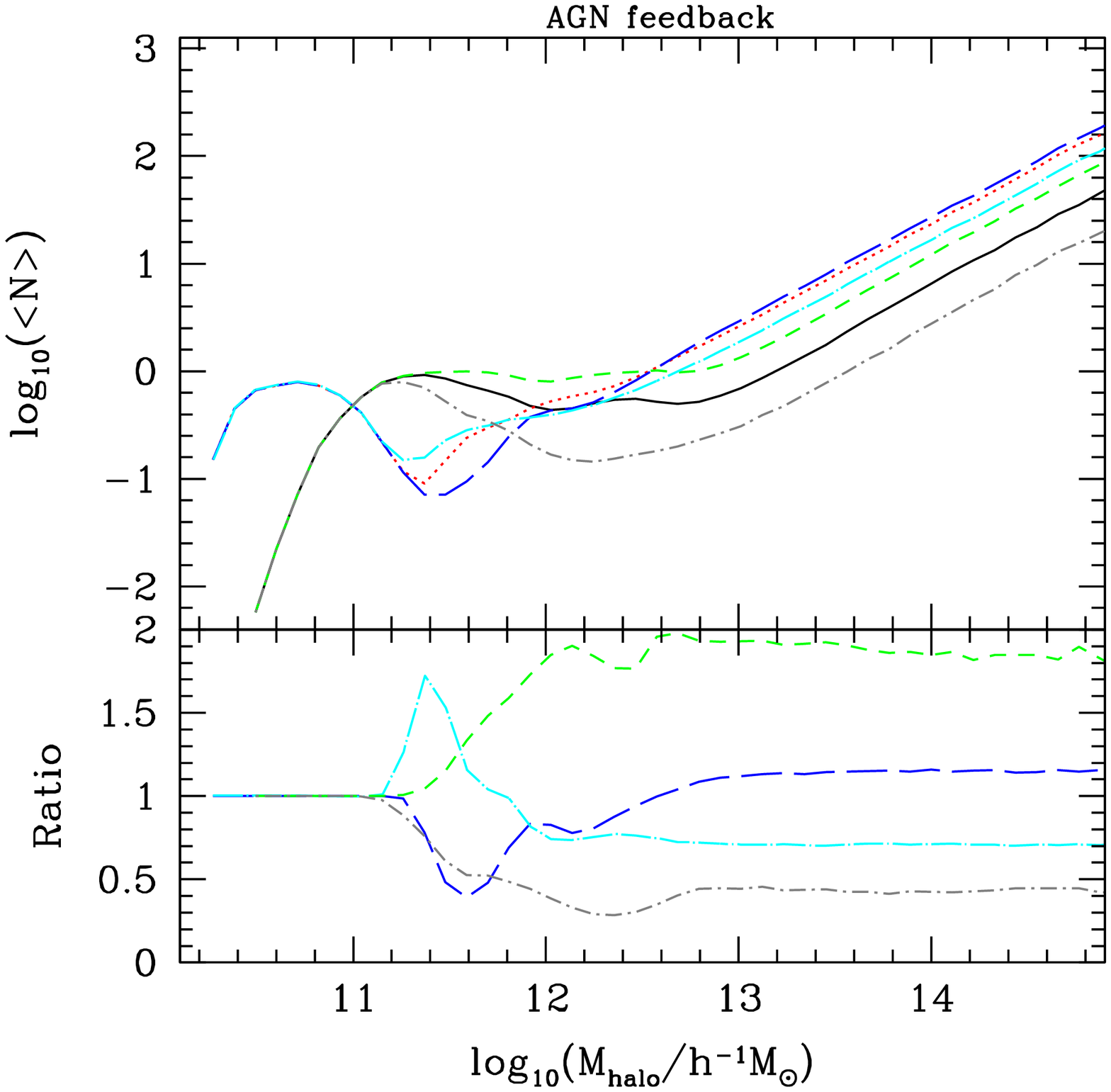}
\includegraphics[width=7.25cm]{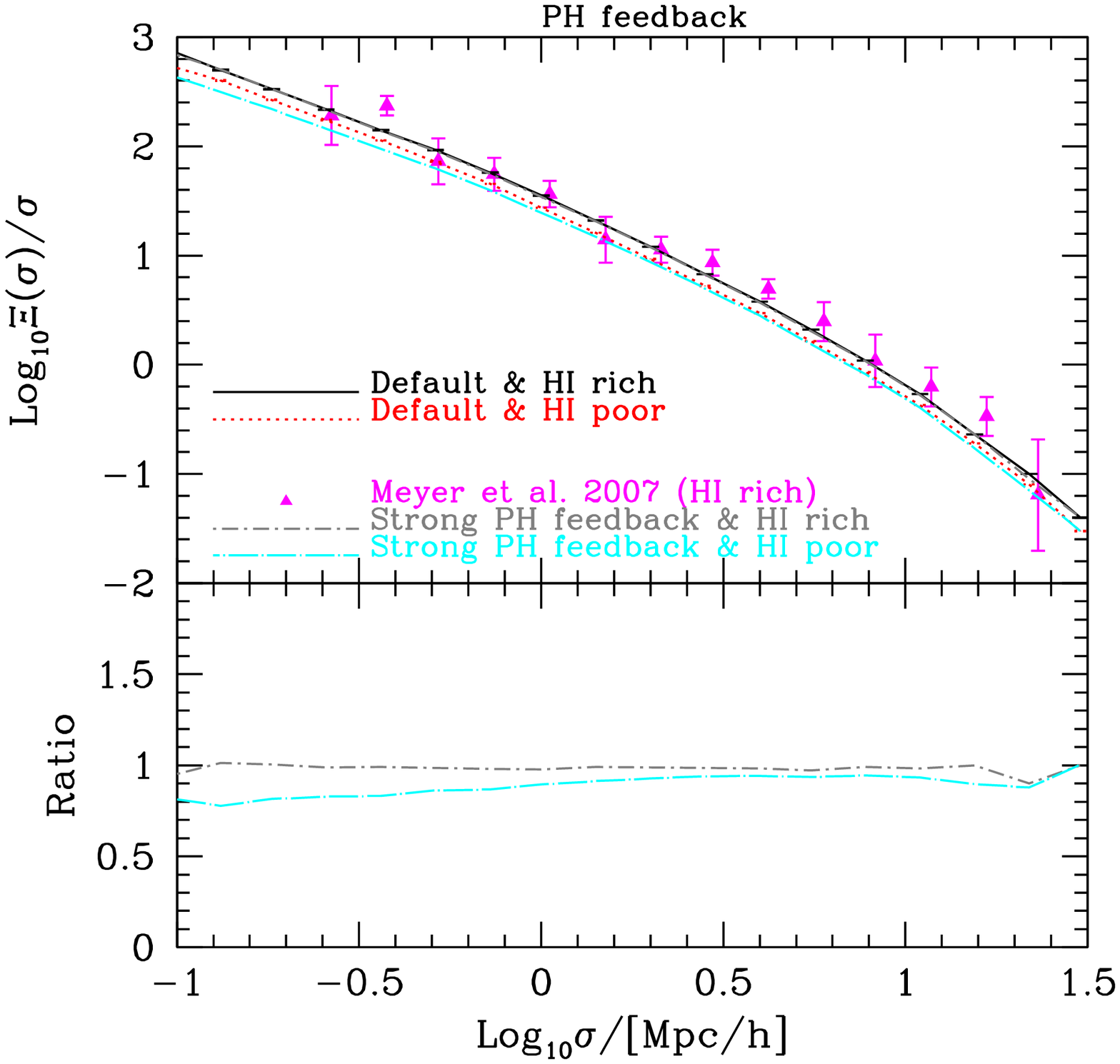}
\includegraphics[width=7.25cm]{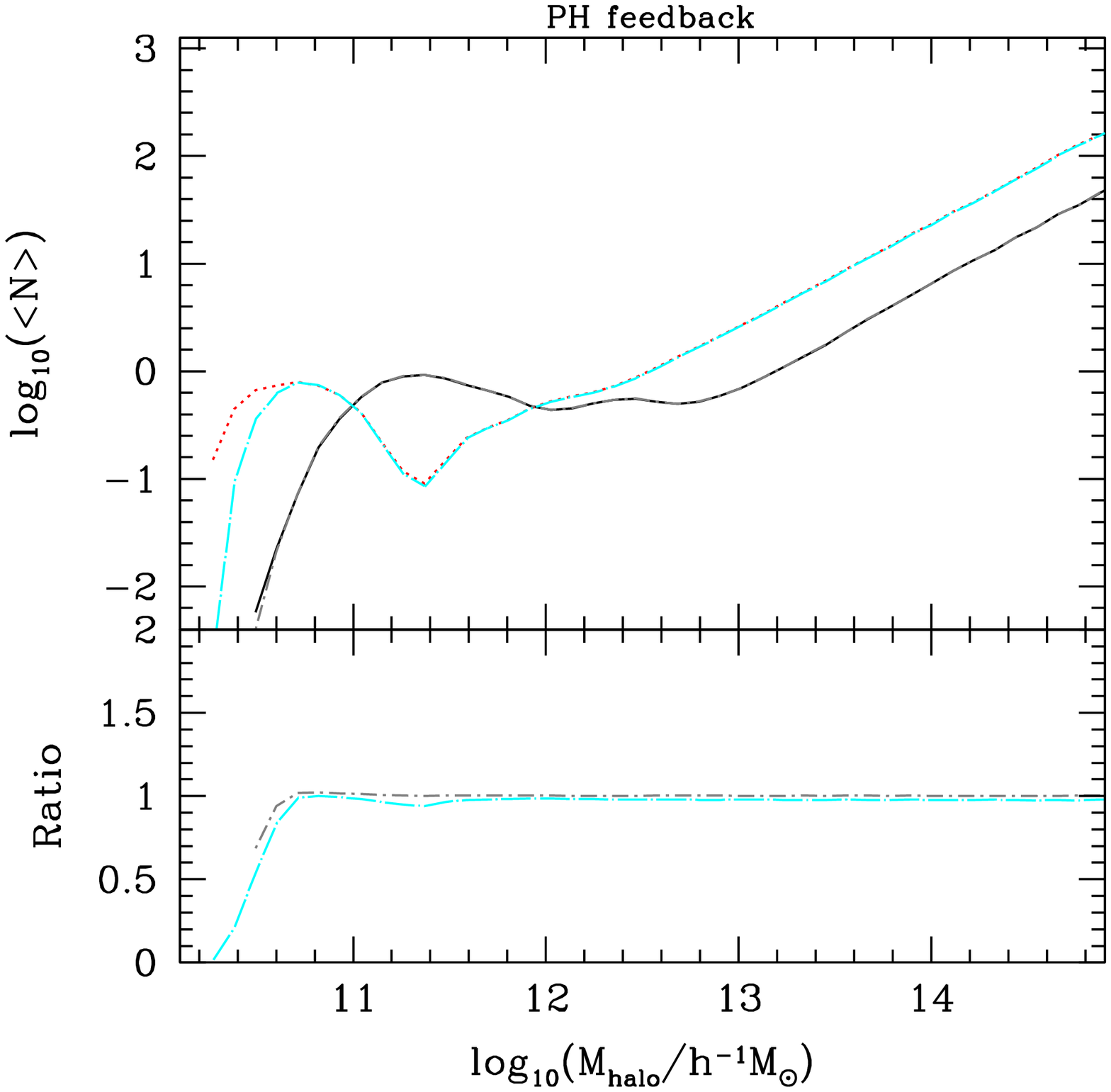}

\caption{The impact of SNe, AGN and photo-ionization feedback (top, middle and
  bottom rows) on the 2-point correlation function (left panels) and the
  halo occupation distribution (right panels). In the upper part of each panel 
  we plot the relevant statistic for HI-poor and HI-rich galaxies, while in the 
  lower part we plot the ratio of the HI-poor or HI-rich statistic with respect to 
  its value in the default (HI-poor or HI-rich) model. Here HI-rich galaxies have masses
  $M_{\rm HI}>10^{9.25}\rm M_{\odot}$ and HI-poor galaxies have HI masses between 
  $10^{8.5} \rm M_{\odot}$ and $10^{9.25} \rm M_{\odot}$. Filled triangles show 
  the 2-point correlation function of gas-rich galaxies obtained by 
  \citet{meyer.etal.2007}, and assorted line types refer to the different
  models, as indicated by the legend in the left panels. See text for 
  further details.
}

\label{SNCF}
\end{figure*}

In Fig~\ref{SNCF} we explore the impact of SNe driven winds (upper panels), 
AGN heating (middle panels) and photo-ionization (lower panels) on the 2-point 
correlation function (left panels) and the HOD (right panels) at $z$=0. In the 
top part of each panel we plot the relevant statistic for galaxies selected 
according to their high or low 
HI mass, while in the bottom part we plot the ratio of the statistic with 
respect to its value in the default model for galaxies of the same HI richness.
Here we define HI-rich galaxies to have $M_{\rm HI}>10^{9.25}h^{-2}\rm M_{\odot}$ 
while HI-poor galaxies
have $10^{8.5}h^{-2}{\rm M_{\odot}}<M_{\rm HI}<10^{9.25}h^{-2}{\rm M_{\odot}}$. For clarity,
we concentrate on the typical weak and strong feedback models (blue and red curves 
respectively) that we considered in Fig~\ref{LFDCOS}. This is necessary because we
expect environmental effects to become apparent when we measure clustering
properties, and we wish to disentangle the effects of environment from 
feedback. We show also the 2-point correlation function of gas rich galaxies 
at $z$=0 measured from the HIPASS survey, presented in \citet{meyer.etal.2007}, and 
the predictions for the 2-point correlation function and the HOD in the default model, 
which is in good agreement with the HIPASS result. 

\paragraph*{Impact of SNe feedback:} The default model 
predicts that HI-rich galaxies (black solid curve) should be more strongly 
clustered than HI-poor galaxies (red dotted curve), which we show in the 
top left panel. Here we see also that strong SNe feedback has negligible 
impact on the clustering strength at large separations of both HI-rich and HI-poor 
galaxies (dotted-dashed and dashed curves respectively), but it influences the 
clustering strength of both HI-rich and HI-poor galaxies at separations 
$\lesssim 1h^{-1}\,\rm Mpc$ (short dashed curve). Reducing the strength of 
SNe feedback leads both HI-poor and HI-rich galaxies to cluster more strongly 
at small separations -- $\lesssim 1 (3) h^{-1}\,\rm Mpc$ for HI-rich (HI-poor) 
galaxies. {Fig.~\ref{SNCF} shows that the clustering amplitude 
of both HI-poor and HI-rich samples becomes similar -- as can be most easily
seen in the lower panel, which shows the ratio of the clustering with respect 
to that of the corresponding HI sample in the default model. This trend can be 
understood by inspecting Figure 3 of \citet{kim.etal.2011}, in which the 
cold gas mass in galaxies is plotted against their host dark matter halo mass. 
Beyond the mass at which AGN heating stops cooling flows, there is a sudden 
break in the cold gas mass - halo mass relation and a dramatic increase in the
scatter. Above this break in mass, the presence of gas is due to accretion of 
halos of lower mass, which have their own gas reservoirs, onto the main halo. 
By reducing the strength of SNe feedback, more gas is able to stay in the cold 
phase in these progenitors and gas is depleted more slowly in the current host.
Because there is a large scatter in galaxy cold gas mass for a given halo mass,
there is a shift in the typical host mass for both HI-poor and HI-rich 
galaxies, leading to an increase in the clustering amplitude.} 

In the top right upper panel we show the predicted HODs as a function of HI richness
(dashed and dotted curves) for all galaxies. Thick (thin) curves indicate strong (weak) 
SNe feedback cases for HI-poor and HI-rich galaxies. As we would expect, stronger SNe 
feedback leads to fewer galaxies when compared to the weaker feedback case, spanning 
the full range of halo masses and regardless of the HI richness of the galaxies. In 
the top right lower panel we plot the ratio of the number of galaxies for the
strong and weak SNe feedback cases compared to the default model. Again this reveals that 
SNe feedback affects the numbers of galaxies regardless of host halo mass, but 
interestingly it shows that the influence of SNe feedback is greater on the numbers of 
HI-rich galaxies compared to HI-poor galaxies.

\paragraph*{Impact of AGN feedback:} In the middle left panel we show that the 
clustering of galaxies selected by their HI richness is largely insensitive to 
the strength of AGN feedback. If AGN feedback is weak, its effect on the clustering 
strength of HI-poor galaxies is negligible and enhances the clustering of HI-rich galaxies
only at small separations. If the AGN feedback is strong, it suppresses the clustering
of both HI-rich and HI-poor galaxies to a similar extent.

As before, we show in the middle right upper and lower panels the predicted 
HODs as a function of HI richness (dashed and dotted curves) for all galaxies, 
where strong (weak) AGN feedback is indicated by thick (thin) curves. The lower panel 
shows that AGN feedback affects halos with masses in excess of $10^{11.2}h^{-1}\rm M_{\odot}$ 
and that it has a greater impact on the numbers of HI-rich galaxies than HI-poor galaxies.

\paragraph*{Impact of Photo-ionization:} Increasing the strength of 
photo-ionization has negligible impact on the clustering strength of HI-rich 
galaxies, but leads to an increase in the clustering strength of HI-poor 
galaxies. This is because increasing the strength of photo-ionization 
corresponds to an increase in the circular velocity scale $V_{\rm cut}$ of
dark matter halos in which galaxies can form. 

We show predicted HODs as a function of HI richness (dashed and dotted curves) 
for all galaxies in the bottom right upper 
and lower panels. Here we consider only the case of strong photo-ionization 
feedback because the typical halo mass affected by weak photo-ionization 
feedback ($V_{\rm cut}$=30 km/s) drops below the resolution limit of the 
Millennium Simulation at early times. Thick lines indicate strong 
photo-ionization feedback whereas thin lines show the default model for 
HI-poor and HI-rich galaxies. It is readily apparent that photo-ionization 
feedback has little influence over the HODs of HI-rich galaxies -- the strong 
feedback case is indistinguishable from the default model. The lower panel 
shows that photo-ionization feedback affects only the lowest mass halos 
($10^{10.7}h^{-1}\rm M_{\odot}$) at $z$=0. 

\section{Summary}
\label{sec:summary}

We have used the version of the {\texttt{GALFORM}} semi-analytical galaxy
formation model \citep[cf.][]{cole.etal.2000} that has been extended by 
\citet{lagos.etal.2011a} to study how feedback from SNe, AGN and 
photo-ionization shapes both the galaxy luminosity and HI mass functions, 
and the spatial clustering of HI galaxies. The advantage of the 
\citet{lagos.etal.2011a} model is that it is one of the first to predict 
explicitly and in a self-consistent manner how much HI should reside in a 
galaxy at a given time \citep[see also][]{fu.etal.2010} and it has 
reproduced successfully the observed HI mass function at low-to-intermediate
masses \citep[cf.][]{lagos.etal.2011b}. The main results of our study can
be summarised as follows;

\vspace{0.4cm}

\noindent$\bullet$ Feedback from SNe regulates the amplitude and 
shape of both the galaxy luminosity function and the HI mass function. The
effect is systematic, with more (less) efficient SNe feedback leading to fewer
(more) galaxies of a given luminosity or HI mass. It also regulates the 
amplitude of the 2-point correlation function of HI-rich galaxies at small 
separations and the numbers of galaxies per halo mass as quantified by the HOD,
regardless of their HI richness.

\vspace{0.2cm}

\noindent$\bullet$ Feedback from AGN has little effect on either the 
faint-to-intermediate luminosity end of the galaxy luminosity function 
or low-to-intermediate mass end of the HI mass function -- its impact is 
strongest for the brightest galaxies, as previous studies have argued 
\citep[e.g.][]{croton.etal.2006,bower.etal.2006}, but it also affects 
the number of highest HI mass galaxies. A similar result has been
reported by \citet{fabello.etal.2011}, and it also helps to explain the cold gas
mass-halo mass relation presented in Fig. 3 of \citet{kim.etal.2011} for the
\citet{bower.etal.2006} model. In contrast, its impact on the clustering of 
HI galaxies is minor or negligible.

\vspace{0.2cm}

\noindent$\bullet$ Feedback from photo-ionization is most pronounced for the 
faintest and most HI-poor galaxies. The redshift at which reionization occurs 
is not important -- but the mass scale, as measured by $V_{\rm cut}$, is, and 
its influence is evident in both the shape and amplitude of the faint end of 
luminosity function and the low-mass end of the HI mass function. Our analysis 
suggests that the low-to-intermediate mass end of HI mass function offers the 
potential to constrain the models of photo-ionization. Interestingly, we find 
that the locally measured HI mass function can constrain the minimum mass of 
dark matter halos that could have hosted galaxies that contributed to 
reionization. This complements existing studies of the high redshift Universe 
that have used, for example, the electron scattering optical depth 
\citep[e.g.][]{choudhury.etal.2008},  the Lyman-$\alpha$ forest and 
\citep[e.g][]{srbinovsky.wyithe.2010} and the UV luminosity function 
\citep[e.g.][]{munoz.loeb.2011} to estimate the minimum halo mass. Our 
analysis suggests that the circular velocities of halos in which galaxy 
formation was suppressed cannot be larger than $\sim\!70\,\rm km/s$.
\vspace{0.2cm}

\noindent$\bullet$ {Strong modes of feedback, as they are implemented in {\texttt{GALFORM}},
act to suppress the clustering strength of galaxies, regardless of their HI-richness. HI-poor
galaxies cluster more strongly than HI-rich galaxies if the strength of SNe feedback is weakened,
whereas there is little change in the clustering of either HI-rich or HI-poor galaxies if the 
strength of AGN feedback is weakened.}

\vspace{0.4cm}

\noindent Our study suggests that forthcoming HI galaxy surveys can be most 
fruitfully exploited scientifically if they are part of a multiwavelength 
campaign. Such campaigns are being planned; for example, DINGO (Deep 
Investigation of Neutral Gas Origins) on ASKAP \citep[cf.][]{meyer.2009} will 
probe the galaxy population out to $z \lesssim 0.4$ and will be combined the 
GAMA (Galaxy And Mass Assembly) survey \citep[cf.][]{driver.etal.2011}.
These data-sets will allow us to study not only the HI properties of galaxies 
that host AGN or young starbursts, but also the effects of environment and 
evolution with redshift.

\section*{Acknowledgements}

HSK is supported by a Super-Science Fellowship from the Australian Research 
Council. CP thanks warmly Danail Obreschkow, Martin Meyer and Aaron Robotham for
helpful discussions. This work was supported by a STFC rolling grant at Durham. The 
calculations for this paper were performed on the ICC Cosmology Machine, 
which is part of the DiRAC Facility jointly funded by the STFC, the Large 
Facilities Capital Fund of BIS, and Durham University.  Part of the research 
presented in this paper was undertaken as part of the Survey Simulation 
Pipeline (SSimPL; {\texttt{http://www.astronomy.swin.edu.au/SSimPL/}). The 
Centre for All-Sky Astrophysics is an Australian Research Council Centre of 
Excellence, funded by grant CE11E0090.

\bibliographystyle{mn2e}

\label{lastpage}

\end{document}